\documentclass{mn2e} 
\usepackage{graphicx}

\def \kms {${\rm{km}\,\rm{s}^{-1}}$}

\def \hg   {H$\gamma$}
\def \hd   {H$\delta$}
\def \hb   {H$\beta$}
\def \ha   {H$\alpha$}
\def \haa   {H$\alpha_{\rm A}$}
\def \haf   {H$\alpha_{\rm F}$}
\def \hgf   {H$\gamma_{\rm F}$}
\def \haaf   {H$\alpha_{\rm A,F}$}

\def \afe  {[$\alpha$/Fe]}

\def \apjs{ApL}
\def \aas{A\&A}

\def \apj{ApJ}
\def \aj{AJ}
\def \mnras{MNRAS}

\title[\ha\ absorption synthesis and the cluster red sequence]{Synthesis of \ha\ absorption in old stellar systems:\\
Formation of the cluster red sequence by ``Downsizing''
}\author[Russell J. Smith]{Russell J. Smith\\
Department of Physics, University of Waterloo, Waterloo, Ontario, Canada \ N2L 3G1 \\ rjsmith@astro.uwaterloo.ca}
\pagerange{\pageref{firstpage}$-$\pageref{lastpage}} \pubyear{2005}
\date{Accepted 21 February 2005}
\begin{document}
\label{firstpage}
\maketitle
\begin{abstract}
We compute population synthesis models for the variation of \ha\ absorption indices (\haa\ and \haf), as a function of age and metallicity in old stellar systems.  The models are based on the STELIB spectral library of J.-F. Le Borgne et al., and defined at a resolution of 3\,\AA\ FWHM. Errors in the age and metallicity responses are derived by bootstrap resampling the input measurements on the stellar library. The indices are found to be highly sensitive to age variation, with only moderate response to metallicity. For galaxies uncontaminated by nebular emission, our \haa\ index is more powerful in breaking the age--metallicity degeneracy than \hb\ or \hgf. Using a sample of red cluster galaxies from Nelan et al., carefully selected to exclude objects with emission, we find a steep decline of \haa\ with velocity dispersion (slope $-0.75\pm0.07$\,\AA\,dex$^{-1}$). The slope can be translated to constraints on age and metallicity scaling relations, incorporating measurement errors and also the model errors determined from the bootstrap method. If the \haa$-\sigma$ slope is due only to age, we obtain Age$\,\propto\sigma^{0.95\pm0.12}$. Because \haa\ depends quite weakly on [Fe/H], a metallicity interpretation would require Fe/H\,$\propto\sigma^{1.2}$ or steeper. The \haa$-\sigma$ slope is consistent with the combined age and metallicity scaling relations reported by Nelan et al. from classical Lick indices. The relations obtained by Thomas et al. significantly under-predict the observed slope.
 The discrepancy could arise from differences in the sample selection. In particular our sample probes a lower mass range, is not explicitly selected on morphological criteria and excludes objects significantly bluer than the red sequence.
 We discuss in detail the effects of emission contamination on the results, and conclude that these are unlikely to yield the observed behaviour in the \ha$-\sigma$ relations. Indeed, similar results are obtained using \haf, despite its different sensitivity to \ha\ and [N\,II] emission lines. The steep age--mass relation supports a ``downsizing'' formation scenario: fainter red-sequence galaxies became quiescent at lower redshifts, $z\la0.5$. This picture accords with recent observations of truncated red sequences in clusters at $z\sim0.7$.
\end{abstract}
\section{Introduction}

In the integrated spectra of old stellar systems, the strength of the Balmer absorption lines reflects the luminosity-weighted effective temperature, which is dominated by the Main Sequence turn-off. In a simple (i.e. single-age, single-metallicity) system, the turn-off luminosity and temperature are sensitive primarily to the age of the population. On this basis, measurements of \hb\ and \hg\ line strengths have been widely used to constrain the ages of early-type galaxies both in clusters and in the field (see Thomas et al. 2004b, and many references therein). Despite extensive work, some fundamental questions remain unresolved, such as the influence of age variations in driving the ``red sequence'' of cluster ellipticals (e.g. Caldwell 2003; Thomas et al. 2004b; Nelan et al. 2005).

The interpretation of the Balmer lines suffers from at least three complications. First, horizontal branch morphology is not included in most stellar population models; if blue Horizontal Branch stars are present, they enhance the Balmer lines and mimic the effect of younger ages (Maraston \& Thomas 2000; Thomas et al. 2004b). Second, the indices are measured on low effective-resolution spectra (usually limited by the internal velocity dispersion of the galaxy itself). This causes the Balmer absorption and neighbouring continua to be blended with a forest of nearby metal absorption lines. Thus rather than true equivalent widths, one measures ``indices'' (such as those of the Lick system, see Burstein et al. 1984), which have residual sensitivity to metallicity and abundance ratios.  Third, many early-type galaxies exhibit weak nebular emission which ``fills in'' the stellar absorption, leading to over-estimates of the population ages (e.g. Gonz\'alez 1993). The contamination is reduced for the higher-order lines, \hg\ and \hd, as a result of the typical line ratios for nebular emission. Unfortunately, however, the usual index definitions for these lines are highly contaminated by metal-line blanketing, and depend more strongly on the overall metallicity (Worthey \& Ottoviani 1997; but see Vazdekis \& Arimoto 1999). Moreover, they are sensitive to the enhancement of $\alpha$-elements, as shown by Thomas, Maraston \& Korn (2004a). By contrast, the low-order line, \hb\ is a less ambiguous tracer of age, if the nebular emission problem can be overcome.

In this paper, we explore whether these trends extend to the lowest-order Balmer line, \ha. In Section~\ref{sec:popsynth}, we review two index definitions, \haa\ and \haf, introduced by Nelan et al. (2005) (Section~\ref{sec:ixdefs}), describe the population synthesis model used (Section~\ref{sec:synthmod}), and determine the sensitivity of the indices to age and metallicity (Section~\ref{sec:synthresults}).
Section~\ref{sec:agemassobs} confronts the synthesis results with measurements for red-sequence galaxies in the NOAO Fundamental Plane Survey (Smith et al. 2004; Nelan et al. 2005). The observed  \ha$-\log\sigma$ relations for a low-emission subsample are derived in Section~\ref{sec:obsdat}, while Section~\ref{sec:agemetvar} discusses the implied age and  metallicity variations. In Section~\ref{sec:booterrs}, we incorporate errors in the population synthesis model, using a bootstrap method. Section~\ref{sec:nebcontam} describes the problem of nebular emission contamination; other potential complications are discussed in Section~\ref{sec:alternative}. On balance, the steep slope of the observed \ha$-\log\sigma$ relations is best reproduced by a strong gradient in age, with low-mass galaxies on average younger than larger objects. We discuss the plausibility of this picture in Section~\ref{sec:discuss}, comparing to observations at intermediate redshifts.
The main conclusions are summarized in Section~\ref{sec:concs}

\section{\ha\ population synthesis}\label{sec:popsynth}

\subsection{Index definitions}\label{sec:ixdefs}

\begin{figure}
\vskip -2mm
\includegraphics[width=85mm,angle=270]{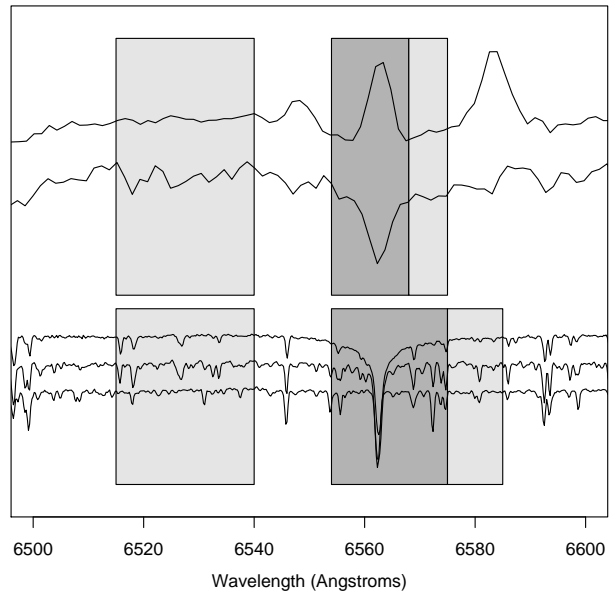}
\vskip -3mm
\caption{\ha\ index definitions from Nelan et al. (2005). The upper shaded boxes indicate the feature and pseudo-continuum pass-bands for \haf; the lower boxes show the pass-bands for \haa.
Comparison spectra show stars from the high-resolution library of Montes, Ramsey \& Welty (1999), with spectral types F8V (top), G8III, K7V. Two galaxies are shown from NFPS, one with emission and the other apparently free from contamination.
}\label{fig:ixdefs}
\end{figure}

\begin{table}
\caption{\ha\ index definitions from Nelan et al. (2005).}\label{tab:ixdefs}
\begin{tabular}{lccc}
\hline
Index & Blue continuum   & Central band & Red continuum\\
\hline
\haa\ & 6515--6540 & 6554--6575 & 6575--6585 \\
\haf\ & 6515--6540 & 6554--6568 & 6568--6575 \\
\hline
\end{tabular}
\end{table}

Throughout this paper, we work with the two index definitions \haa\ and \haf, introduced by Nelan et al. The band-passes defining these indices are reproduced in Table~\ref{tab:ixdefs}, and shown in Figure~\ref{fig:ixdefs}, in comparison to some representative  spectra for stars and galaxies. The two indices were defined with the practical intention of being measurable for as many objects as possible in the NFPS dataset. Within NFPS, only the spectra obtained at the WIYN telescope have sufficient coverage in the red to measure \ha, and even then, only for the most nearby galaxies. The major practical consideration was thus to ensure the red continuum was as close as possible to the line. A second constraint is the presence of two [N\,II] emission lines bracketing \ha, at 6548\,\AA\ and 6583\,\AA. Although it is possible to place the blue pseudo-continuum on the blue side of the 6548\,\AA\ line, pushing the red pseudo-continuum beyond the  6583\,\AA\ line is impractical given the spectral-range constraint. Two alternative definitions were proposed, with different treatments of the red continuum.

The narrower index, \haf, has continuum bands placed to avoid the [N\,II] lines in galaxies with an emission component. Where emission is present, \haf\ will be weakened by infilling, or will become negative if the emission dominates. For pure-absorption objects, the sensitivity of \haf\ is likely to be reduced because the absorption spreads into the continuum band, especially at large velocity dispersion. The wider definition, \haa\ has continuum bands better separated from \ha\ itself, and so provides a better indicator of stellar absorption. Where emission is present, both the central band and the red continuum band are contaminated. The net effect will depend on the ratio of [N\,II] / \ha; given the LINER-like ratios typically observed in ellipticals (Phillips et al. 1986), the \ha\ emission is likely partially compensated by [N\,II] in most cases.

For age determination, a key issue is to separate the hydrogen line itself from the surrounding  metallic lines.  In the \ha\ region, the density of metal lines is much lower than for the blue Balmer lines. One potential contaminant arises from Ca\,I at 6573\,\AA, which falls in the central pass band of \haa\ and in the red continuum of \haf. For emission-free galaxies, the treatment of this feature is likely the most important difference between the two indices. The abundance of Ca in elliptical galaxies
seems to track Fe, rather than following Mg as would be  expected for an $\alpha$-element (e.g. Thomas, Maraston \& Bender 2003b; Cenarro et al. 2004, and references therein). Thus although Ca\,I\,6573\,\AA\ may introduce [Fe/H] contamination, this should not depend strongly on \afe.

In conclusion, our \haa\ and \haf\ indices are practically motivated, and were not tuned for optimal performance in population synthesis models.
However, as we show in the following sections, the NFPS \ha\ indices appear to be excellent age indicators, with very little residual sensitivity to the metallicity.

\subsection{Population synthesis}\label{sec:synthmod}

To predict the \haa\ and \haf\ index values for simple stellar populations, we first determine their responses to physical properties of stars, then combine the contributions using stellar properties along a given evolutionary isochrone, weighted by an assumed initial mass function.

For this analysis, we use the STELIB spectral library of Le Borgne et al. (2003), which covers a metallicity range of
$-2.0\la$[Fe/H]$\la+0.5$ at a nominal spectral resolution of 3\AA. This library is well-matched to the NFPS native resolution (also $\sim$3\AA). The \haa\ and \haf\ indices were measured on the STELIB library spectra using the program {\sc indexf} (Cenarro et al. 2001), as used also for the NFPS galaxy measurements.

\begin{figure}
\includegraphics[width=85mm]{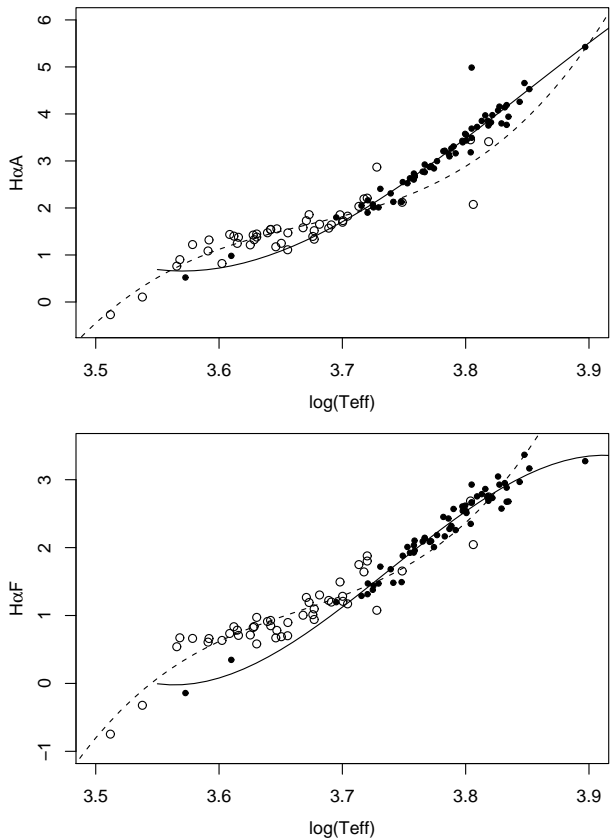}
\vskip -3mm
\caption{Temperature dependence of the \haa\ and \haf\ indices, as determined from the STELIB library stars. Dwarf stars ($\log\,g>3.4$) are plotted with filled symbols and solid line; giant stars with open symbols and dashed line. The index values have been corrected to [Fe/H]=0 and $\log\,g=3.4$, to isolate the temperature terms. Each fit is a cubic in $\log\,T_{\rm eff}$.}\label{fig:fitfnsteff}
\end{figure}

\begin{figure}
\includegraphics[width=85mm]{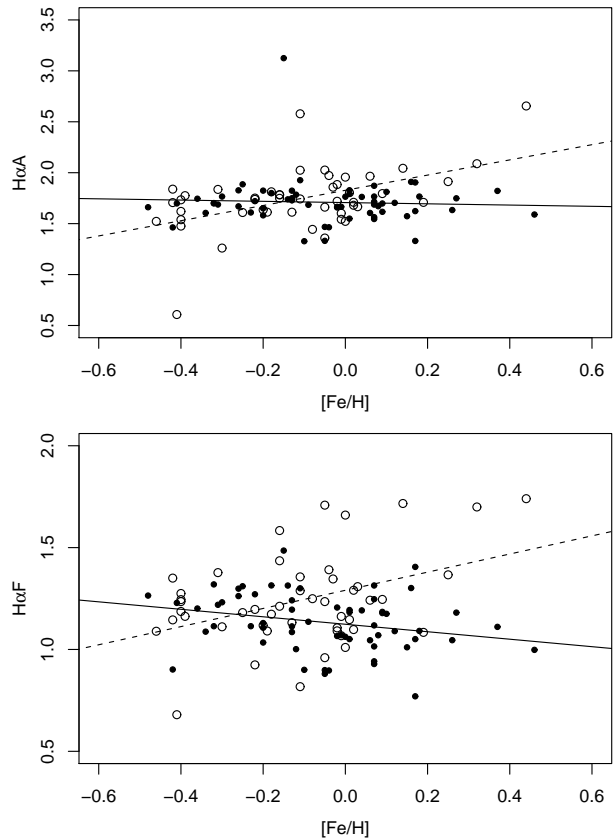}
\vskip -3mm
\caption{Metallicity dependence of the \haa\ and \haf\ indices, as determined from the STELIB library stars. Symbols and line-styles as in Figure~\ref{fig:fitfnsteff}. The index values have been corrected to $\log\,T_{\rm eff}$=3.7 and
$\log\,g=3.4$, to isolate the metallicity term. The fits are linear in [Fe/H].}\label{fig:fitfnsmet}
\end{figure}

\begin{figure}
\includegraphics[width=85mm]{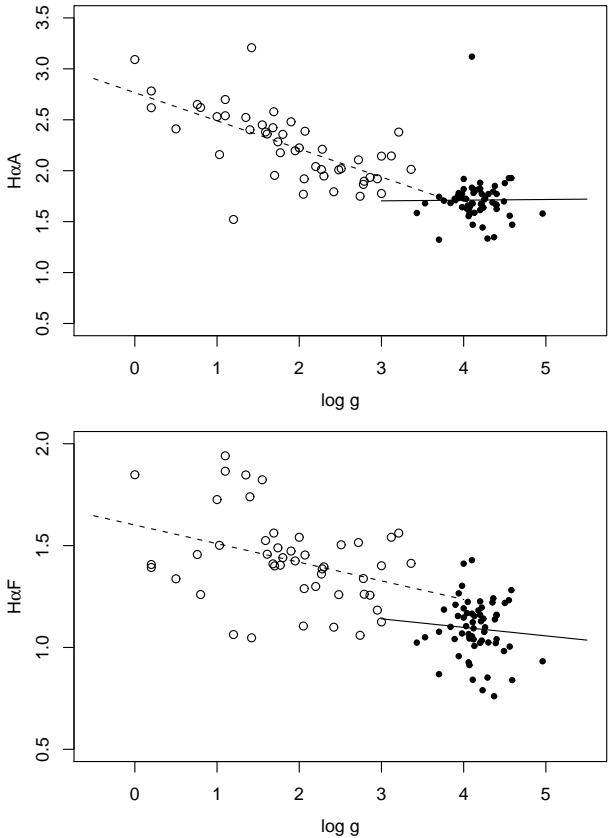}
\vskip -3mm
\caption{Gravity dependence of the \haa\ and \haf\ indices, as determined from the STELIB library stars. Symbols and line-styles as in Figure~\ref{fig:fitfnsteff}. The index values have been corrected to  $\log\,T_{\rm eff}$=3.7 and [Fe/H]=0, to isolate the gravity term. The fits are linear in $\log\,g$. Note that because gravity and temperature are strongly correlated for giants, the real variations in \ha\ at fixed temperature and metallicity are much smaller than the total range shown here.}\label{fig:fitfnsgrav}
\end{figure}

We derive separate fitting functions for the dwarf and giant sequences, separated at $\log\,g=3.4$ (Figures~\ref{fig:fitfnsteff}--\ref{fig:fitfnsgrav}). Since the indices are especially sensitive to temperature, we model them with a cubic in $\log\,T_{\rm eff}$, and linear correction terms in $\log\,g$ and [Fe/H] (alternative parametrizations of the $T_{\rm eff}$ dependence have been tested, without significant effect on our conclusions). The stars allowed to contribute to the fits have temperatures below 7000\,K (for 47 giant stars) or 9000\,K (for 61 dwarfs) and metallicity [Fe/H]$>-0.5$. The stellar parameter range covered by the library and the fits limits the validity of our models to ages $>$1\,Gyr and metallicities --0.5$\la$[Fe/H]$\la$+0.3. At metallicities above twice-solar, the library coverage is very sparse, compromising predictions for the most enriched systems such as truly giant ellipticals. Normal early-type galaxies, with $\sigma\la200$\,\kms\ exhibit only modestly super-solar values, ([Fe/H]$\la+0.3$) when measured from Fe-dominated features, rather than $\alpha$-enhanced lines.

To determine the \haa\ and \haf\ that would be measured on an integrated population of a single age and single metallicity, we sum the contributions from different stellar masses along the appropriate isochrone. For these calculations, we adopt the Padova theoretical tracks, taken from Salasnich et al. (2000),  with solar-scaled abundance ratios and metallicities $Z = 0.008, 0.019, 0.040, 0.070$. The  contribution of each mass interval is weighted by the number of stars in the interval, assuming a Salpeter initial mass function (IMF)  $N(M)dM\propto\,M^{-2.35}$, and by the R-band luminosity of the star, as a proxy for the \ha\ continuum flux.

\subsection{Results}\label{sec:synthresults}

\begin{figure}
\includegraphics[width=85mm]{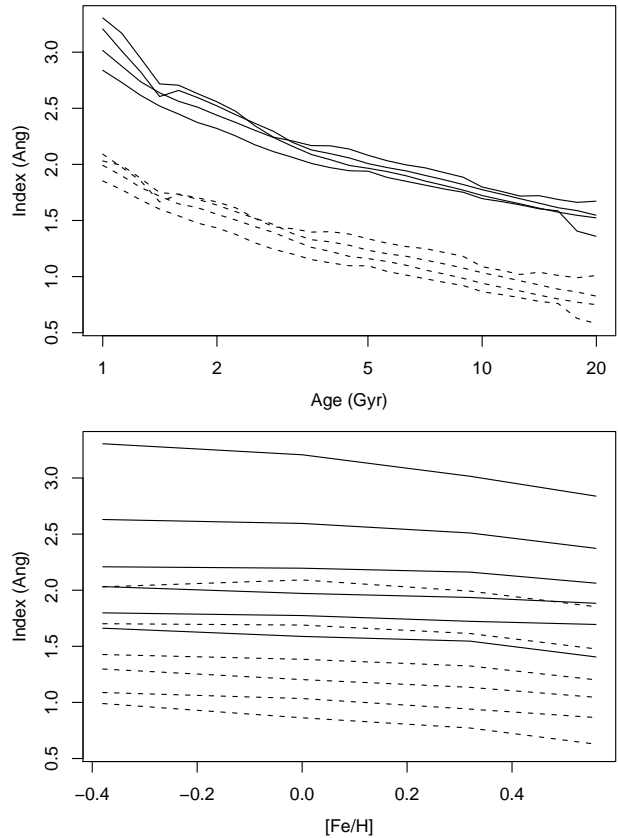}
\vskip -3mm
\caption{Index predictions for \haa\ (solid lines) and \haf\ (dashed lines).
The upper panel shows the indices as a function of age, for constant
metallicity $Z$=0.008 (highest), 0.019, 0.040 and 0.070. The lower panel shows index variations
with metallicity, at constant age of 1.00 (highest), 1.78, 3.16, 5.62, 10.0, 17.8 Gyr.}\label{fig:synth}
\end{figure}

Figure~\ref{fig:synth} shows the synthetic \haa\ and \haf\ as a function of input age and metallicity. As usual with such models, relative changes in the indices are expected to be robust than the their absolute values. In particular, we can extract the following ``responses'' to the population parameters:
\[
\frac{\partial{\rm H}\alpha_{\rm A}}{\partial\log\,Age} = -0.783_{-0.031}^{+0.100} \ \ \ ; \ \ \
\frac{\partial{\rm H}\alpha_{\rm A}}{\partial\,{\rm [Fe/H]}} = -0.161_{-0.215}^{+0.150}
\]
\[
\frac{\partial{\rm H}\alpha_{\rm F}}{\partial\log\,Age} = -0.683_{-0.029}^{+0.084}\ \ \ ; \ \ \
\frac{\partial{\rm H}\alpha_{\rm F}}{\partial\,{\rm [Fe/H]}} = -0.278_{-0.158}^{+0.086}
\]
In practice these responses are derived from a simultaneous linear fit to \haaf(log Age, [Fe/H]), over the subset of models with Age$>$3\,Gyr. This fit reproduces the index predictions with an rms scatter of less than $0.025$\,\AA.
The errors in the responses are estimated through a bootstrap algorithm, described below.

\begin{figure}
\includegraphics[width=85mm]{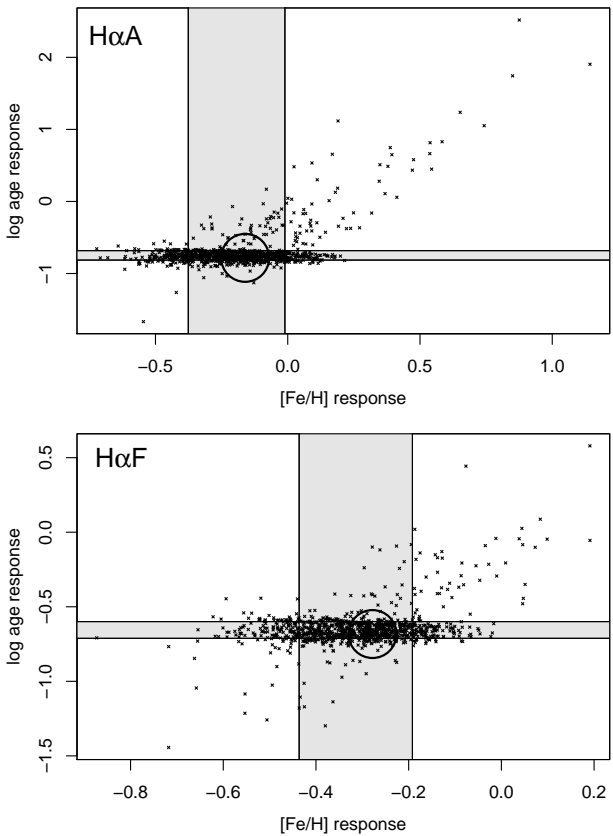}
\vskip -3mm
\caption{Parameter responses for the two \ha\ indices. The large circles indicate the best-fit values derived from the original synthesis; small crosses show responses generated by bootstrap resampling the input stellar library. Grey regions show the 68.3\% (i.e. $1\sigma$) intervals for each parameter separately.}\label{fig:bootresp}
\end{figure}

Physically, the age dependence of \ha\ absorption arises primarily from the  increased temperature, hence stronger Balmer absorption, at the Main Sequence turn-off in younger populations. Our results are contrary to the claim by Caldwell et al. (2003), that \ha\ should be insensitive to age because the R-band luminosity is dominated by the red giant branch, in which \ha\ absorption is weak and independent of age.  The underlying Padova/Salpeter population model instead yields roughly equal flux-contributions from the giants and the dwarfs. The ratio of \ha\ contributions for the youngest models is 25:75 (giant:dwarf), while the oldest models have around 50:50 ratios.

The error estimates are derived from 1000 bootstrap realisations of the input stellar library generated by resampling (with replacement) from the initial input list.
The aim is to propagate the uncertainties in the fitting functions, in particular due to sparse coverage of the stellar atmosphere parameter space by STELIB. Implicitly, the method assumes that coverage of the parameter space is incomplete, but is not systematically biased.
For each realisation, new fitting functions are computed (according to the same criteria used above), and used to predict index values for the grid of stellar populations. The resulting distribution for the parameter responses is shown in Figure~\ref{fig:bootresp}. Although most realisations yield results close to the default model, some $\sim$\,3\% lead to dramatically different parameter responses, with \ha\ absorption {\it increasing} with age. These cases arise when the two cool dwarf stars are both absent from the resampled library. In such cases, the cubic
$\log{}T_{\rm eff}$ term is badly under-constrained, yielding unphysically large \ha\ contributions from the lower Main Sequence. This does not imply a problem in the default synthesis, since in this case, the library does adequately constrain the cubic fit.  The ``failure'' of some bootstrap trials illustrates an important aspect of the bootstrap results, that each realisation samples the parameter space less well than the original library. Our error estimates are therefore conservative in this respect. In Section~\ref{sec:agemetvar}, the bootstrap results will be further propagated to constrain the age and metallicity scaling relations from the observed \ha$-\sigma$ relations.

Our models are constructed on the assumption of solar-scaled chemical abundances, while in fact elliptical galaxies typically have \afe=0.1--0.3 (e.g. Kuntschner et al. 2001). Thomas et al. (2003a, 2004a) have generalised linestrength prediction models to allow for non-solar abundance ratios. These models involve corrections based on stellar theoretical atmosphere calculations (e.g. Tripicco \& Bell 1995) for the contribution of various metallic absorption lines. They find that \hd\ and \hg\ are highly sensitive to \afe-variations, since the hydrogen lines are embedded in a forest of metal lines which contribute to the measured index. By contrast, the \hb\ line lies in a spectral region with a lower density of metal lines, and is almost insensitive to \afe. As discussed in Section~\ref{sec:ixdefs}, high-resolution stellar spectra reveal very clean continuum regions around \ha\ (Figure~\ref{fig:ixdefs}). The Ca\,I 6573\,\AA\ absorption line likely does not introduce strong \afe\ dependence, given the anomalous behaviour of Ca relative to other $\alpha$-elements. Thus, although we cannot readily determine the \afe-response of the \ha\ indices defined here, it seems unlikely that \afe-variations seriously compromise relative age measurements using \haa.

Taking the ratio of the parameter responses, we find that \haa\ is $f\approx4.8$ times more sensitive to a decade change in age, than to a decade change in metallicity, while for \haf\ the equivalent ratio is $f\approx1.9$. For comparison, in the Thomas et al. (2003a, 2004a) models, the standard Lick \hb\ index has $f\approx1.7$, while the \hg$_{\rm A,F}$ and \hd$_{\rm A,F}$ indices all have $f\approx1.0$. The bootstrap simulations reveal substantial uncertainty in the metallicity response, and hence $f$, so it is premature to conclude that \haa\ offers a dramatic improvement over the more widely used features. However, the two \ha\ indices, and especially \haa, appear promising in this respect, and worthy of future study to constrain further the metallicity dependence.

\begin{table*}
\caption{Predictions for \haa\ and \haf\ indices as a function of age and metallicity. Each line represents a constant metallicity track; each column a track of constant log(age), with age in Gyr, as indicated by the column headings.}
\vskip -1mm
\begin{tabular}{llcccccccccccccc}
\hline
& [Fe/H] & 9.0 &  9.1&  9.2&  9.3&  9.4&  9.5&  9.6&  9.7&  9.8&  9.9&  10.0 &  10.1&  10.2&  10.3\\
\hline
\haa
& $-0.38$  & 3.305  &2.945  &2.705  &2.559  &2.346  &2.209  &2.165  &2.080  &1.995  &1.927  &1.798  &1.717  &1.687  &1.672  \\
& $\phantom{+}0.00$  & 3.207  &2.824  &2.659  &2.523  &2.368  &2.196  &2.095  &2.006  &1.942  &1.863  &1.774  &1.694  &1.614  &1.547  \\
& $+0.32$  & 3.015  &2.736  &2.562  &2.439  &2.305  &2.161  &2.042  &1.965  &1.898  &1.812  &1.723  &1.648  &1.575  &1.524  \\
& +0.56  & 2.838  &2.615  &2.448  &2.321  &2.176  &2.063  &1.971  &1.939  &1.849  &1.781  &1.695  &1.639  &1.586  &1.359  \\
\hline
\haf
& $-0.38$  & 2.030  &1.876  &1.735  &1.666  &1.521  &1.427  &1.400  &1.336  &1.268  &1.215  &1.088  &1.019  &1.011  &1.009  \\
& $\phantom{+}0.00$  & 2.092  &1.853  &1.732  &1.638  &1.522  &1.384  &1.309  &1.234  &1.181  &1.113  &1.035  &0.963  &0.889  &0.826  \\
& +0.32  & 1.992  &1.794  &1.648  &1.556  &1.447  &1.325  &1.225  &1.160  &1.101  &1.023  &0.940  &0.872  &0.801  &0.750  \\
& +0.56  & 1.852  &1.682  &1.541  &1.434  &1.303  &1.201  &1.122  &1.096  &1.013  &0.949  &0.865  &0.812  &0.760  &0.582  \\
\hline
\end{tabular}
\end{table*}

\section{A strong age--mass relation along the red sequence?}\label{sec:agemassobs}

\subsection{Comparison to NFPS data}\label{sec:obsdat}

In this section, we compare the predicted \ha\ absorption to observed values reported by Nelan et al. (2005) from the NFPS.
We use the ``full-resolution'' ($\sim$3\,\AA\ FWHM) measurements, corrected for aperture and velocity-broadening effects. The NFPS galaxy sample was selected by apparent magnitude ($R_{\rm Tot}<17$) and colour ($\Delta(B-R)>-0.2$), relative to the red-sequence ``ridge'' in each cluster (Smith et al. 2004). There is no cut on the red side of the ridge. Note especially that we do not explicitly select by galaxy morphology; thus some of the galaxies included in this sample would not be present in a hand-picked set of bona fide ellipticals and S0s.
Moreover, the NFPS sample explicitly selects against any massive ellipticals with relatively blue colours.
It is for this reason that we refer always to red-sequence galaxies, rather than early-type galaxies. Although the full NFPS database includes linestrength data for $\sim$4000 galaxies, \ha\ is beyond the range of most of the spectra. Only 689 confirmed cluster members, spanning a redshift range $cz=3600-12700$\,\kms, have measurements for \haa, \haf, plus necessary auxiliary data (velocity dispersions and emission-line estimates). The redshift range of the \ha\ dataset is less than the median depth of NFPS (15000\,\kms). Because the NFPS galaxy sample is flux limited, and \ha\ is observed for only the nearby objects, the typical mass of the \ha\ sample is slightly fainter than that of the whole survey.

As discussed in the Introduction, a major concern with \ha\ is that the age-sensitive stellar absorption may be contaminated by nebular emission, both at \ha\ itself, and in the neighbouring [N\,II] lines. It is critical therefore to work with an subsample of galaxies in which nebular contamination is minimized. Nelan et al. (2005) determined emission equivalent widths for the [O\,III] line at 5007\,\AA, and for \hb, by fitting and dividing an appropriate stellar continuum. For the present analysis, we exclude 250 galaxies with emission at \hb\ and/or at [O\,III], based on a 1$\sigma$ detection limit. This rejection scheme is much stricter than the fixed thresholds imposed by Nelan et al. To ensure sufficient data quality, we impose a further selection for \haa\ errors less than 0.3\,\AA. After applying these cuts, the sample comprises 410 galaxies.

\begin{figure}
\includegraphics[width=85mm]{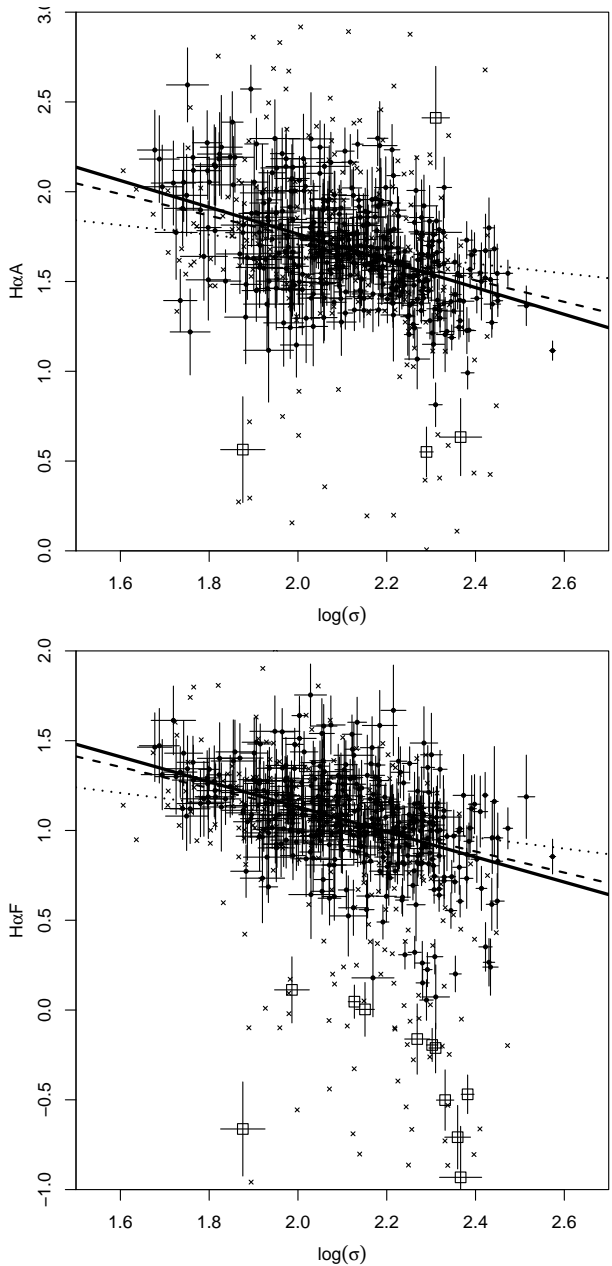}
\vskip -3mm
\caption{Observed \ha$-\log\sigma$ relations from NFPS data of Nelan et al. (2005).
Galaxies from the low-emission subsample (see text), with \haa\ errors less than 0.3\,\AA, are shown as solid points with error bars. Open squares mark galaxies rejected from the final fit (solid line) by an iterative 3-$\sigma$ clipping process. Small crosses indicate galaxies failing either the low-emission selection or the low-error selection (or both). In each panel, the dashed lines shows the slope predicted for \ha$-\log\sigma$ using the age and metallicity scaling relations derived by Nelan et al. from traditional Lick indices. The scaling relations of Thomas et al. (2004b) result in predicted slopes shown by the dotted line.}
\label{fig:data}
\end{figure}

The \ha$-\sigma$ relations for this low-emission subsample are shown in Figure~\ref{fig:data}. The data are fit using an iterative 3$\sigma$ clipping, with the effect of rejecting a further four outliers in \haa\ and eleven in \haf.
For the median $\sigma\approx125$\,\kms, the fitted values are \haa$\approx$1.69\,\AA\ and \haf$\approx$1.06\,\AA. Taking both data and models at face value, these indicate average ages of 10--13\,Gyr (\haa) or 7--9\,Gyr (\haf), if $0.0<$[Fe/H]$<0.3$. Potential causes of this discrepancy are discussed in Sections~\ref{sec:nebcontam} and \ref{sec:alternative}.

Differential observations, e.g. the change in \ha\ absorption with galaxy mass, are expected on general grounds to be more robust than absolute values. The observed \haa$-\sigma$ relation has slope of $-0.75\pm0.07$\,\AA\ per decade in $\sigma$. For \haf\ the equivalent slope is $-0.70\pm0.07$. In the remainder of this section, we attempt to interpret these slopes first in terms of age and metallicity variations, and then consider alternative explanations.

\subsection{Age or metallicity variations}\label{sec:agemetvar}

Given the measured gradient of \haa$-\sigma$ (or \haf$-\sigma$) and a pair of values summarizing the index response to age and [Fe/H], we can constrain the exponents ($\alpha,\beta$) of scaling relations such that Age$\,\propto\,\sigma^\alpha$ and Fe/H$\,\propto\,\sigma^\beta$. Naturally, since only one gradient is measured, the viable values for $\alpha$ and $\beta$ fall along a degenerate linear track, along which the metallicity and age effects combine to produce the observed \ha\ slope.
By setting each of ($\alpha,\beta$) to zero in turn, we can determine the age and metallicity trends required if only one of these parameters varies with $\sigma$.

If we naively adopt simply the ``best'' parameter responses from the population synthesis of Section~\ref{sec:popsynth}, and allow only for the errors in the observed \haa$-\sigma$ relation we would require either Age$\,\propto\sigma^{0.95\pm0.09}$
(if metallicity is constant) or Fe/H$\,\propto\sigma^{4.6\pm0.5}$ (if age is constant).
From \haf\ the equivalent slopes are ($\alpha,\beta$)=($1.02,2.5$) with comparable errors.
The metallicity trend required is far in excess of determinations using Lick indices (e.g. Kuntschner et al. 2001, who similarly constrained their fits by assuming no age variation, reporting Fe/H\,$\propto\sigma^{\approx1.0}$). The age interpretation implies a factor of five in age over the 60--300\,\kms\ range in velocity dispersion, which is also surprisingly strong (but see Section~\ref{sec:discuss}).

A similarly strong age--mass relation has been claimed by Nelan et al. (2005), based on traditional Lick indices from NFPS. Their reported scaling relations are Age$\,\propto\sigma^{0.67\pm0.15}$ and [Fe/H]$\,\propto\sigma^{0.47\pm0.05}$. Allowing {\it both} parameters to vary according to the Nelan et al. relations, we would expect an \haa\ slope of $-0.52$\,\AA\ from age and an additional $-0.08$\,\AA\ from [Fe/H] variations, totalling $-0.60\pm0.12$\,\AA\ per dex, marginally consistent with the observed $-0.75\pm0.07$.
The Nelan et al. analysis is independent of the present work, in the sense that it does not employ \ha\ absorption, and uses a much larger sample of $\sim$3500 galaxies.
However, the sample characteristics are similar, since both works are based on the same survey.

An independent study of 124 galaxies by Thomas et al. (2004b), also based on traditional Lick indices, and analysed using the same models as Nelan et al., yields Age$\,\propto\,\sigma^{\approx0.25}$. The Thomas et al. scalings predict a \haa\ slope of $-0.27$\,\AA\ ($-0.18$\,\AA\ from age and $-0.09$\,\AA\ from metallicity),  apparently quite incompatible with our observed relation. Compared to the sample used here, the Thomas et al. compilation is weighted towards more massive objects ($\sigma\ga$100\,kms, with median of $\sim$200\,\kms), and also to fairly clean E or S0 morphologies.
Both the mass-range and the morphological mix could influence the results obtained, as could the explicit colour-magnitude selection. Nelan et al. commented on an apparent steepening of the relation for low-mass systems, which could account for some of the disagreement with Thomas et al., and comparable studies. Moreover,
there are indications that NFPS galaxies with stronger disks show a steeper age--mass relation than pure ellipticals or bulge-dominated S0s.
Both effects act in the sense which could explain the discrepancy between NFPS and Thomas et al.
Finally, the NFPS colour--magnitude selection excludes any massive `blue ellipticals'. If included, and if present in sufficient numbers, such objects would tend to flatten the age--mass relation.

Caldwell et al. (2003) studied a sample of 175 early-type galaxies with $\sigma=50-300$\,\kms, similar to the range probed in this paper. Using their favoured index combination, Caldwell et al. derive a steep age--mass slope of Age$\,\propto\sigma^{\approx1.0}$ (see their Figure 21 and Table 9), in agreement with Nelan et al. and with our \ha\ results. Again, this seems argue for a dependence on the mass-range, with giant ellipticals favouring a flatter age--mass relation, and low-$\sigma$ objects yielding a steeper slope.

\subsection{Effect of synthesis model errors}\label{sec:booterrs}

As already hinted at, the above discussion neglects an important source of error, arising from uncertainties in the population synthesis model itself. These errors can be included in the constraints by propagating the bootstrap-response distribution of Figure~\ref{fig:bootresp}. Recall that each bootstrap-realisation of the input stellar library was used to determine a new estimate of the parameter responses. Most of the bootstrap realisations yield responses close to the ``best'' values, and hence yield ($\alpha,\beta$)-constraints similar to the picture described in the previous section. On the other hand, some bootstrap points show strong [Fe/H] responses (e.g. around -0.5\,\AA\ in \haa\ per dex Fe/H). These clearly improve the chances of observing a strong \haa$-\sigma$ trend with neither a strong age-trend nor an excessive [Fe/H]$-\sigma$ scaling. Finally, the spray of ``anomalous'' bootstrap results, due to realisations which lack cool dwarfs, assigns some (low) probability to quite unexpected ($\alpha,\beta$) pairs.

\begin{figure}
\includegraphics[width=85mm]{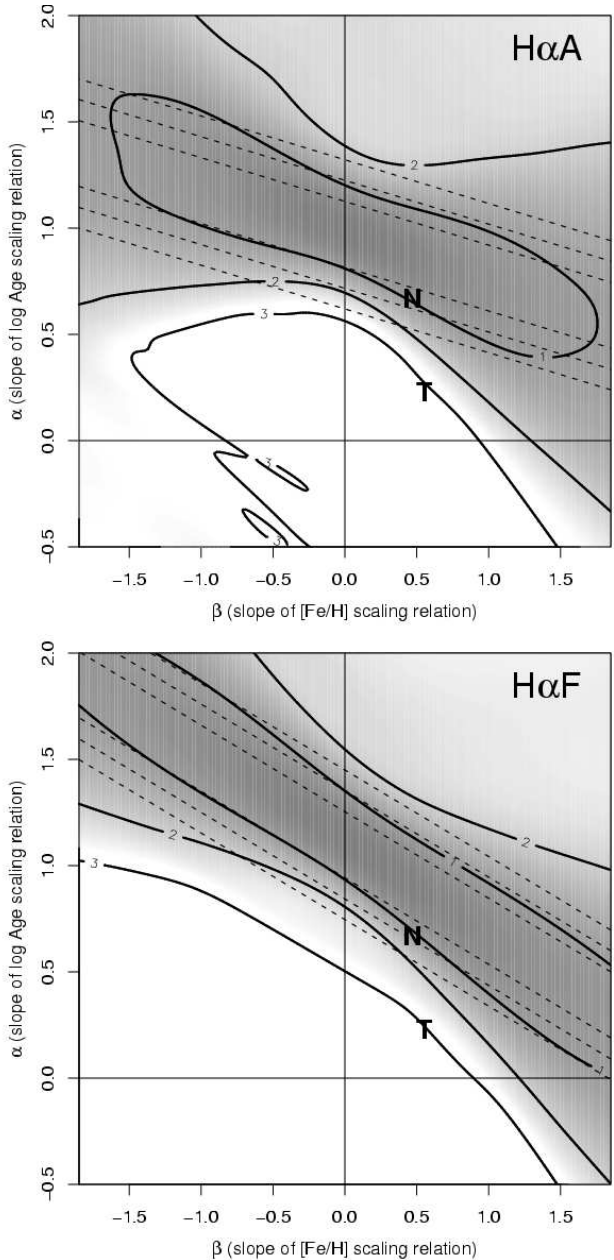}
\vskip -3mm
\caption{The upper panel shows $\mathcal{L}(\alpha,\beta)$, the likelihood of the observed \haa\ gradient, given scaling relations Age$\,\propto\sigma^\alpha$ and Fe/H$\,\propto\sigma^\beta$. In addition to the observational errors on the \haa$-\sigma$ slope, the likelihood explicitly accounts for synthesis model errors, using the bootstrap simulations. Bold contours show the resulting 1$\sigma$, 2$\sigma$ and 3$\sigma$ confidence regions for two parameters. For comparison, dashed contours show the equivalent limits if model errors are neglected. The labels `N' and `T' refer to the scaling relations from Nelan et al. (2005) and Thomas et al. (2004b) respectively. The lower panel shows the equivalent results for \haf.}\label{fig:bootscal}
\end{figure}

Combining these simulations, Figure~\ref{fig:bootscal} shows the likelihood function $\mathcal{L}(\alpha,\beta)$ allowing for the bootstrap errors and  also for measurement error in the observed \ha$-\sigma$ slopes. As expected, the range of viable scaling relations is considerably expanded when the model errors are included. Formally, for a constant-metallicity sequence, we obtain Age$\,\propto\sigma^{0.99\pm0.12}$ from \haa. In the constant-age case, the Fe/H$-\sigma$ relation has a slope of $\beta>1.5$ at 1$\sigma$ ($\beta>1.2$ at 2$\sigma$), with no meaningful upper limits. Thus if a constant age is imposed, the implied metallicity trend remains much steeper than values favoured by Lick-index studies. Allowing both parameters to vary, the Nelan et al. (2005) scalings lie at the $1\sigma$ contour. The scalings of Thomas et al. (2004b) remain disfavoured at the $\sim$3$\sigma$ level.

Two curious aspects of Figure~\ref{fig:bootscal} are worthy of brief comment. First, the change in shape of the likelihood contours, which now form a closed curve at 1$\sigma$ suggests that allowing for the bootstrap errors has actually tightened the constraint on the metallicity-gradient taken separately. However, if we integrate over the age-gradient as a nuisance parameter, to obtain the one-parameter constraint, the  vertical ``flaring'' of the likelihood function compensates for this effect.
Second, because the likelihoods were derived from only 1000 bootstrap realisations, the 3$\sigma$ contour is not well-determined. The noisy features at the lower-left of the \haa\ panel result from the anomalous bootstrap solutions, i.e. from the upper right of Figure~\ref{fig:bootresp}. The 3$\sigma$ contour is retained in the plot mainly to show this effect, and to illustrate the discrete tracks which contribute to $\mathcal{L}(\alpha,\beta)$.

\subsection{Nebular emission contamination}\label{sec:nebcontam}

The greatest obstacle to the use of \ha\ absorption as a stellar age indicator is its potential for contamination by nebular emission, both at \ha\ itself, and in the neighbouring [N\,II] lines. In Section~\ref{sec:obsdat}, we used the emission measurements at \hb\ and [O\,III] to reject the objects most strongly affected. This approach  lacks precision for very low-level contamination, since the \hb\ emission is weaker than \ha. Moreover, the \hb\ measurements are themselves made in the presence of imperfectly-known stellar absorption, potentially leading to underestimated equivalent widths.

As a result, our ``low-emission'' subsample will include some galaxies with low-level contamination. In general, weak emission will yield smaller values for \haf, causing galaxies to appear older. In the lower panel of Figure~\ref{fig:data}, the scatter of outliers below the mean \haf$-\sigma$ relation suggests such an effect, with a few galaxies showing negative \haf\ values, i.e. net emission. For \haa\ the situation is more complicated, since [N\,II] in the red continuum region leads to increased index values, which may explain the absence of such dramatic outliers in the \haa$-\sigma$ relation. This compensating influence depends on the ratio of [N\,II] to \ha\ emission, which is strongly correlated with absolute magnitude (Phillips et al. 1986). Most of the outliers in \haf\ are fairly high mass objects, where the [N\,II] compensation will be most efficient. In this case however, it is surprising that they are not detected at [O\,III] which should also be strong in such objects.

Although clipping of deviant points can remove a small number of outliers, a more serious concern is whether the mean relation itself might be driven by variations in emission characteristics, rather than age or metallicity. One test for this is to repeat the fits, allowing for terms in \hb\ and [O\,III] emission, as well as velocity dispersion. In these fits the \haa$-\sigma$ and \haf$-\sigma$ slopes are very little changed, and the emission-line coefficients are not significant. To generate the observed \ha$-\sigma$ trends with emission variations would require increasing \ha\ emission with increasing mass (opposite to the behaviour at \hb).
The similarity of results from \haa\ and \haf\ is also difficult to explain if the \ha$-\sigma$ slopes are mainly due to emission: because the two indices respond differently to [N\,II], we would need either very low [N\,II]/\ha\ ratio throughout, or else very little variation in this ratio as a function of mass. Either option is contrary to the behaviour observed in galaxies with strong emission (Phillips et al. 1986). Moreover, the amount of emission would have to be very precisely determined at any given mass, in order to retain a tight \ha$-\sigma$ relation at all.

A related concern is whether the \hb-based selection itself biases the sample against low-mass weak-absorption objects. This is conceivable, since separating emission and absorption components of \hb\ is more difficult at low-$\sigma$ (Nelan et al. 2005). The signature of such a bias would be an increased proportion of emission-rejected galaxies below the fit, at small $\sigma$. The effect is not readily apparent in Figure~\ref{fig:data}. Running the fits with a variety of alternative emission rejection schemes, including that of Nelan et al., the \haa\ gradients are stable at the 15\% level. For \haf, the results are less stable, yielding steeper slopes when the looser selection criteria are used.

Finally, emission does not easily account for the absolute age discrepancy between \haa\ and \haf. Recall that \haa\ yields older ages at given $\sigma$ than \haf. The emission effect should cause both indices to yield older ages, but \haa\ should be less strongly affected, as a result of the [N\,II] compensation.

In conclusion, emission contamination remains a very serious concern for \ha\ absorption studies. Emission undoubtedly generates outliers from the observed relations, some of which cannot easily be excluded using other observations. On balance however, it appears unlikely that emission variations are responsible for the strong, decreasing trends of \haa\ and \haf\ with velocity dispersion.

\subsection{Alternative explanations}\label{sec:alternative}

As well as sensitivity to nebular emission, our result is subject to a number of other caveats, relating to assumptions and limitations of the synthesis model. The following paragraphs consider three such concerns: Blue Horizontal Branch (BHB) stars, non-solar abundance ratios and IMF slope variation.

The Balmer lines measure the luminosity-weighted temperature of a stellar population, so the \haa\ trend could be generated by any hot stellar component which becomes stronger at low $\sigma$. Maraston \& Thomas (2000) and Thomas et al. (2004b) have explored models with BHB populations, and shown that the strong \hb\ absorption of some ellipticals might indeed arise from BHB stars rather than from younger ages. Because these stars have blue continua, they contribute more strongly to the higher-order lines than to \hb\ (Schiavon et al. 2004), and their impact will be weakened further at \ha. Quantitative modelling of the relative responses should offer one approach to constraining the importance of BHB stars in old populations.

We have argued that non-solar abundance ratios are unlikely to have a strong effect on \ha, since there are no strong metallic features within the band-passes used by our indices. The only feature of concern is the Ca\,I 6573\,\AA\ line, which affects the \haa\ and \haf\ indices in opposite senses. The similarity of the scaling relation slopes obtained from the two indices suggests that variations of Ca abundance with $\sigma$ do not strongly influence the results. On the other hand, the difference in absolute ages implied by the two indices could be related to this line. In particular, if Ca is under-abundant for the whole sample (relative to STELIB), then \haa\ will be anomalously low (leading to older ages), while \haf\ will be larger (yielding younger ages). Thus the sense of the effect is as observed. However, to force \haa\ and \haf\ to give consistent ages at given $\sigma$, we need a large metallicity difference, of $\ga$0.3\,dex. On balance, simple differences in flux-calibration or resolution-matching seem more likely explanations for the 0.1\,\AA\ offset required to resolve the absolute age inconsistency.

Finally, we have explored the effects of
 changing the assumed IMF slope in the synthesis. For power-law exponents in the range 1.0--3.5 (cf. default 2.35), the predictions for \haa\ and \haf\ vary by $\la$0.3\,\AA. Thus the full range of $\sim$0.75\,\AA\ cannot be generated by plausible levels of variation in the IMF slope.

\section{Discussion}\label{sec:discuss}

Having reviewed some of the potential systematic effects, in this section we consider whether a steep age--mass relation is compatible with other constraints, focusing on observations of cluster galaxies at lookback times of a few Gyr. In what follows, we will adopt the \haa-derived relation of Age\,$\propto\sigma^{0.95\pm0.12}$, which assumes no metallicity variations. Considering that Lick index studies suggest a mix of age and metallicity effects, this slope should be regarded as an upper limit.

\begin{figure}
\includegraphics[width=85mm,angle=270]{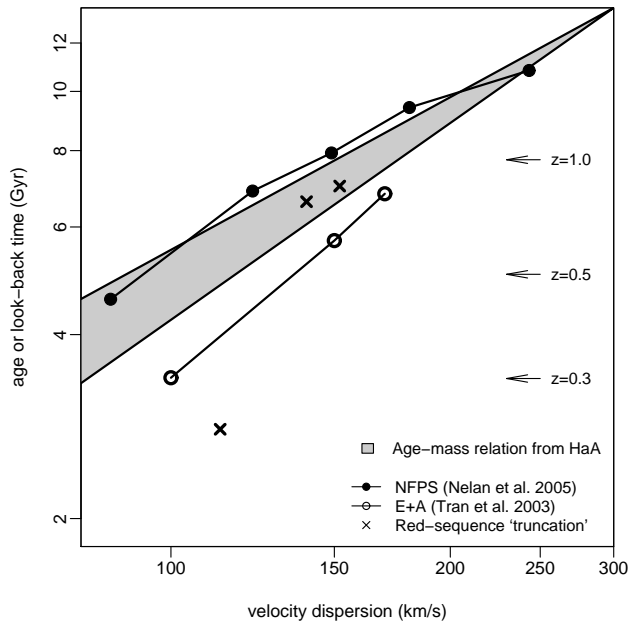}
\caption{Formation history of red-sequence cluster galaxies. The shaded region shows the mean age--mass relation Age\,$\propto\sigma^{0.95\pm0.12}$, as suggested by the \haa$-\sigma$ relation of this paper. Plausible metallicity variations could flatten this relation by 10--40\% (see Figure~\ref{fig:bootscal}). The ages are calibrated by fixing the high-mass galaxies to the age of the universe, 13.7\,Gyr.
The results of Nelan et al. (2005), using multiple Lick indices for $\sim$3000 NFPS galaxies are shown as the filled circles. Open circles show the characteristic velocity dispersion of E+A galaxies in clusters from z=0.3 to z=0.8 (Tran et al. 2003). These results may be compared to the low-redshift age measurements if the E+A objects are interpreted as newly-quiescent objects fading onto the red sequence. Finally, crosses indicate the mass scale corresponding to faint-end truncation of the red sequence in clusters at z=0.24 (Smail et al. 1998), z=0.75 (de Lucia et al. 2004) and z=0.83 (Goto et al. 2005).}\label{fig:tform}
\end{figure}

Since the absolute ages implied by the \ha\ synthesis are unacceptably old, we fix the age zero-point by forcing the most massive galaxies to the age of the universe. (Note that if \haf\ were used instead, this rescaling would not be necessary, and all other results would be essentially unchanged.) In this section we adopt the WMAP cosmological parameters, $(h,\Omega_{\rm m},\Omega_\Lambda)=(0.71,0.27,0.73)$ (Bennet et al. 2003), yielding a maximum age 13.7\,Gyr. In this case, the age--mass relation implies a mean age of only 3\,Gyr for the smallest objects, i.e. with $\sigma\sim60$\,\kms, while the observed scatter around the \haa$-\sigma$ relation allows for an intrinsic age range of 2--5\,Gyr for these galaxies. The age--mass relation is shown in Figure~\ref{fig:tform}, together with some of the other results discussed below. In all cases, the ``age'' or ``formation redshift'' is to be understood as
a luminosity-weighted mean age for the stars and does not imply the galaxy assembly time. At times earlier than this ``age'', the galaxy may have been present in the cluster but with the bluer colours indicative of a star-forming system.

The steep \haa$-\sigma$ trend then appears to favour a picture in which galaxies now at the faint end of the red sequence ceased forming stars only at modest redshifts, $z\la0.3$. A testable implication is that the red sequence in high redshift clusters should be deficient in faint galaxies, with the depletion mass increasing with increasing redshift.
Recent observations of $z\sim$0.8 clusters have indeed suggested that the red sequences are depleted at the faint end (de Lucia et al. 2004; Goto et al. 2005). Converting their observed magnitudes to a velocity dispersion (after accounting for k-correction and passive evolution), the faintest non-star-forming objects in these clusters have $\sigma\sim150$\,\kms, at a look-back time of $\sim$7\,Gyr. For comparison, the age--mass relation derived in this paper also places the current age of such galaxies at $\sim$7\,Gyr. Intriguingly, there is evidence for red-sequence truncation at much smaller lookback times. Smail et al. (1998) observed clusters at $z\approx0.24$ (lookback time of only $\sim$3\,Gyr); their Figure 5 shows a marked cut-off in the red sequence, at a magnitude corresponding to $\sigma\sim$110\,\kms. For comparison, our age--mass relation suggests ages of $\sim$5\,Gyr for such objects. A rigorous comparison of these works is not trivial, since they do not use consistent definitions for the ``truncation magnitude''. However, there is very suggestive agreement with our \haa\ results, both in terms of the absolute age of the faint red sequence galaxies, and in the trend for fainter objects joining the red sequence at lower redshifts.

Additional support comes from the rapidly-evolving characteristic mass scale of E+A (or post-starburst) galaxies, as determined by Tran et al. (2003). Such galaxies can be interpreted as objects which have just ceased forming stars, and are in the process of fading onto the red sequence. The results of Tran et al. suggest that E+A galaxies observed at $z\sim0.8$ (lookback time 7\,Gyr) will fade to $\sigma\sim170$\,\kms\ red-sequence objects today, while those in the E+A phase at $z\sim0.3$ (3.5\,Gyr) will fall onto the red sequence at $\sigma\sim100$\,\kms. Again, there is a suggestive level of agreement with our \haa\ results: the slope of the E+A mass evolution itself suggests Age$\propto\sigma^{1.3}$, slightly steeper than our result.
The E+A ``ages'' fall $\sim$1\,Gyr younger than our relation would predict. Qualitatively this is  expected, since the E+A phase signals the last star-formation episode, while our relation describes the mean stellar age, which is necessarily older.

In conclusion, local galaxy studies which suggest rapid recent evolution in the red sequence population (Caldwell et al. 2003; Nelan et al. 2005; this work) are compatible with direct observations of galaxy populations at intermediate redshifts. It is not, of course, implied that the faint red galaxies were themselves assembled at such recent epochs, but rather that star formation ceased at that time, perhaps following expulsion of remaining gas, or its rapid consumption in a starburst. On the other hand, it may be a challenge to reconcile such recent star formation with the tightness of the colour-magnitude relation itself (e.g. Bower, Lucey \& Ellis 1992), without invoking age--metallicity ``conspiracy'' models.

\section{Conclusions}\label{sec:concs}

We have investigated the behaviour of \ha\ absorption indices in single-age, single-metallicity stellar populations, and compared to observed trends in red-sequence cluster galaxies. The principal conclusions are:

\begin{enumerate}

\item
The \ha\ absorption strength, measured using \haa\ or \haf, appears promising as a probe of stellar ages in integrated spectra, with only weak response to metallicity variations.
The indices are expected to be more robust against non-solar [$\alpha$/Fe] ratios than the high-order Balmer lines.

\item
The principal drawback to the \ha\ method is its sensitivity to nebular emission contamination, which can only be minimized through careful sample selection.

\item
At face value, the slope of the observed \haa$-\sigma$ relation, from NFPS, requires a factor of $\sim$5 in age over the observed $0.7$\,dex range in  velocity dispersion, with younger ages for lower-mass objects. A pure metallicity interpretation requires $\Delta$[Fe/H]$>1.1$\,dex over the same range, inconsistent with other constraints.

\item
The combined, more modest, age and metallicity trends reported by Nelan et al. (2005) for NFPS reproduce the observed \haa$-\sigma$ slope within the errors. By contrast, the scaling relations of Thomas et al. (2004b) significantly under-predict the slope.

\item
If the \ha$-\sigma$ slope is due purely to age, the derived age--mass relation implies ages of only a few Gyr for low-luminosity red-sequence galaxies in clusters. Interpreted as the time at which such galaxies ceased forming stars, the ages broadly agree with an observed deficit of faint red-sequence members in clusters at $z\approx0.7$.

\end{enumerate}

In summary, \ha\ absorption is a potentially powerful, but neglected, probe of early-type galaxy formation epochs. Future work with improved stellar libraries should aim to constrain further the metallicity dependence of \ha\ indices, and determine optimal passbands for index definitions. The analysis in this paper supports a steep age--mass relation along the red sequence of cluster galaxies. Although the most massive cluster ellipticals are indeed very old, our work suggests that the faint red-sequence galaxies became quiescent as recently as $z\la$0.5. Qualitatively, this ``downsizing'' scenario (cf. Cowie et al. 1996) agrees with recent detections of a faint-end depletion of the red sequence at higher redshifts, and with the evolution of E+A galaxy masses over the past $\sim$7\,Gyr.

\section*{Acknowledgments}
It is a pleasure to thank Mike Hudson for helpful and encouraging comments on this work, and all of my collaborators in NFPS, for their help in assembling the dataset used here.
An anonymous referee provided several useful and constructive suggestions. All of the analysis was conducted, and all of the figures generated, using ``R'', an open-source language and environment for statistical computing (http://www.R-project.org).

{}

\label{lastpage}
\end{document}